\documentclass[twocolumn,pre,showpacs,aps,nofootinbib]{revtex4}

\usepackage{graphicx}
\graphicspath{{./fig/}}

\usepackage{bm}

\newcommand{\kk}{\bm{k}}
\newcommand{\rr}{\bm{r}}
\newcommand{\uu}{\bm{u}}
\newcommand{\xx}{\bm{x}}
\newcommand{\dd}{\mathrm{d}}
\newcommand{\Eq}[1]{Eq.~(\ref{#1})}
\newcommand{\Fig}[1]{Fig.~\ref{#1}}

\newcommand{\yjour}[4]{, #2 {\bf #3}, #4 (#1).}
\newcommand{\yjfm}[3]{, J. Fluid Mech. {\bf #2}, #3 (#1).}
\newcommand{\ypf}[3]{, Phys. Fluids {\bf #2}, #3 (#1).}
\newcommand{\ypp}[3]{, Phys. Plasmas {\bf #2}, #3 (#1).}
\newcommand{\ypr}[3]{, Phys.\ Rev.\ {\bf #2}, #3 (#1).}
\newcommand{\yprl}[3]{, Phys.\ Rev.\ Lett.\ {\bf #2}, #3 (#1).}
\newcommand{\yapj}[3]{, Astrophys. J. {\bf #2}, #3 (#1).}

\begin{document}
\preprint{NORDITA 2003-11 AP}

\title{Bottleneck effect in three-dimensional turbulence simulations}

\author{Wolfgang Dobler}
  \email{Wolfgang.Dobler@kis.uni-freiburg.de}
  \affiliation{Kiepenheuer-Institut f\"ur Sonnenphysik,
  Sch\"oneckstra\ss{}e 6, D-79104 Freiburg, Germany}
\author{Nils Erland L.\ Haugen}
  \email{Nils.Haugen@phys.ntnu.no}
  \affiliation{Department of Physics, The Norwegian University of Science
  and Technology, H{\o}yskoleringen 5, N-7034 Trondheim, Norway}
\author{Tarek A.\ Yousef}
  \email{Tarek.Yousef@mtf.ntnu.no}
  \affiliation{Fluid Technology Group,
  Faculty of Engineering Science and Technology,
  NTNU,
  Kolbj{\o }rn Hejes vei 2B, N-7491 Trondheim, Norway}
\author{Axel Brandenburg}
  \email{Brandenb@nordita.dk}
  \affiliation{NORDITA, Blegdamsvej 17, DK-2100 Copenhagen \O, Denmark}

\date{\today}

\begin{abstract}
At numerical resolutions around $512^3$ and above, three-dimensional
energy spectra from turbulence simulations begin to show noticeably
shallower spectra than $k^{-5/3}$ near the dissipation
wavenumber (`bottleneck effect').
This effect is shown to be significantly weaker
in one-dimensional spectra such as those obtained in wind tunnel turbulence.
The difference can be understood in terms of
the transformation between one-dimensional
and three-dimensional energy spectra under the assumption that the
turbulent velocity field is isotropic.
Transversal and longitudinal energy spectra are similar and can both
accurately be computed from the full three-dimensional spectra.
Second-order structure functions are less susceptible to the bottleneck
effect and may be better suited for inferring the scaling exponent from
numerical simulation data.
\end{abstract}

\maketitle

\section{Introduction}

Based on dimensional analysis, Kolmogorov \cite{Kolmogorov41} concluded that
the energy spectrum for isotropic hydrodynamic turbulence has the form
\begin{equation}
  E(k) = C_{\rm K} \epsilon^{2/3}k^{-5/3}
  \label{Kolmogorov}
\end{equation}
for wave numbers $k$ in the inertial range between the energy-carrying
and the dissipation wave number $k_{\rm d}=\epsilon^{1/4}\nu^{-3/4}$
(where $\nu$ denotes the kinematic viscosity, $\epsilon$ the spectral
energy flux, and $C_{\rm K}$ is nowadays called `Kolmogorov constant').
This scaling has been confirmed experimentally over several orders
of magnitude \cite{Champagne1978,Sreenivasan95}.
Nevertheless, numerical simulations consistently show excess power just
before the dissipation wavenumber $k_{\rm d}$, which manifests itself
particularly at
high resolution \cite{simulations_bottleneck}.
This phenomenon has been named `bottleneck effect'
\cite{bottleneck,LohseMuellerG} and
is usually explained by the lack of smaller-scale vortices at wave numbers
$k > k_{\rm d}$, which makes the energy cascade less efficient around
$k_{\rm d}$.
According to a related interpretation, the bottleneck effect is the
consequence of viscosity stabilizing small vortex tubes against the kink
instability \cite{WPEAB1995}.

The effect is particularly strong when an unphysical hyperviscosity is
used; see Refs~\cite{BSC98,BM00} for results from two-dimensional
hydromagnetic turbulence.
In experimental data, on the other hand, the bottleneck effect is
less pronounced \cite{SJ93,SaddoughiVeeravalli1994},
and it has previously been noticed that it is much
weaker when one-dimensional energy spectra are used \cite{BM00}.
In the present paper we discuss a simple relation between one-dimensional
and three-dimensional energy spectra, which agrees well with the
simulation data and explains this difference.

\begin{figure}
  \centering
  \includegraphics[width=0.48\textwidth]{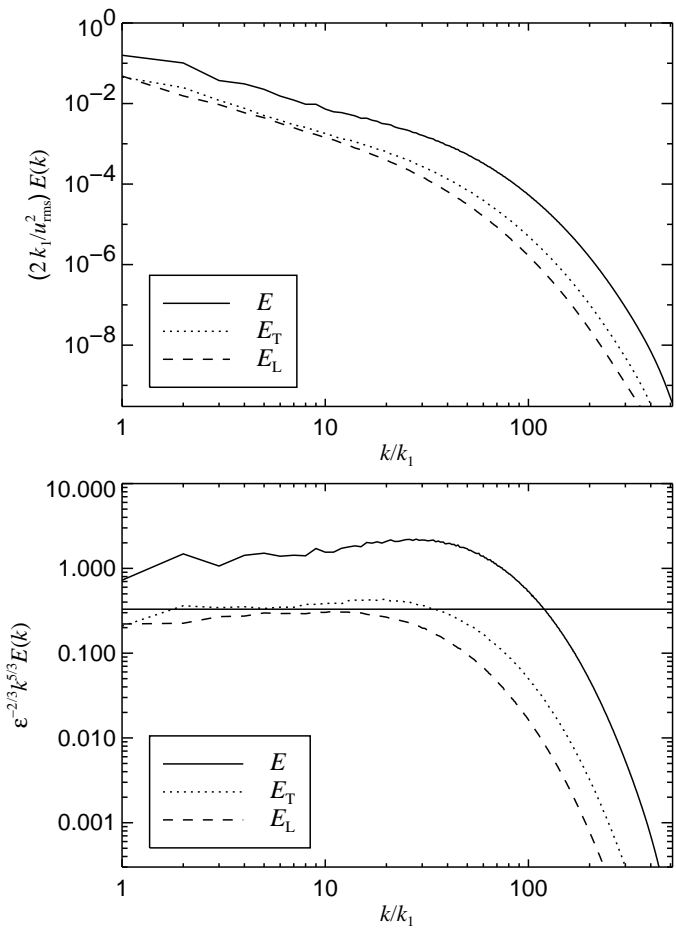}
  \caption{Comparison of the averaged one-dimensional longitudinal and
    transversal spectra, $E_{\rm L}(k)$ and $E_{\rm T}(k)$, respectively,
    with the three-dimensional spectrum $E(k)$ for a forced turbulence
    simulation at $1024^3$ grid points.
    Top: the spectra.
    Bottom: `compensated spectra'
    $\epsilon^{-2/3}k^{5/3}E(k)$,
    $\epsilon^{-2/3}k^{5/3}E_{\rm L/T}(k)$; the
    horizontal line represents a $k^{-5/3}$ Kolmogorov spectrum.
    The local maximum of $E(k)$ around $k\approx30$ represents the
    bottleneck effect.
    The dissipation wave number is
    $k_{\rm d} = \epsilon^{1/4}\nu^{-3/4}\approx 200$.
  }\label{pspec_bottleneck_512}
\end{figure}

The data we use for discussing the bottleneck effect are from
a weakly compressible isothermal
three-dimensional forced turbulence simulation at a numerical resolution
of $1024^3$ grid points.
The forcing has vanishing net helicity and the forcing wave number
$k_{\rm f}$ is between 1 and 2.
The box size is $L_x = L_y = L_z = 2\pi$, which discretizes the wave
numbers in units of $k_1=1$.
The viscosity $\nu$ is chosen such that the Reynolds number based on
the inverse mean forcing wave number, $u_{\rm rms}/(\nu\overline{k}_{\rm f})$,
is around $1700$.
The Taylor microscale is $\sqrt{5}u_{\rm rms}/\omega_{\rm rms} \approx 0.14$,
where $\omega_{\rm rms}$ is the root mean square vorticity,
so the corresponding Taylor microscale Reynolds number is $350$.
The average dissipation rate $\epsilon$ is such that $k_{\rm d}/k_{\rm f}$
is around $130$.
The root mean square Mach number is between $0.17$ and $0.20$; for this 
type of weakly compressible simulations, we find that the energies of
solenoidal and potential components of the flow have a ratio
$E_{\rm pot}/E_{\rm sol} \approx 10^{-4}\mbox{--}10^{-2}$ for most
scales; only towards the Nyquist frequency the ratio increases to about $0.1$.
Even for Mach numbers between $0.5$ and $10$, this ratio is only about $0.1$
to $0.2$ \cite{PorterWoodward1998,BoldyrevNordlundPadoan2002}.
It is thus reasonable to assume that compressibility is irrelevant for the
results presented here.
This is also supported by the fact that incompressible pseudo-spectral
simulations at a resolution of $1024^3$ show a
bottleneck effect \cite{simulations_bottleneck}.

The simulations discussed here were carried out using a high-order
finite-difference code \cite{PencilCode}
and thus complement the results that have so far been obtained
using spectral codes.
Figure \ref{pspec_bottleneck_512} shows the
three-dimensional spectrum $E(k)$ together with the longitudinal and
transversal one-dimensional spectra $E_{\rm L}(k)$ and $E_{\rm T}(k)$ in
the upper panel.
The `compensated spectra' in the lower panel allow easy identification of
the bottleneck effect and show that it is practically present only in the
three-dimensional spectrum $E(k)$.

In order to reduce the otherwise huge fluctuations,
the one-dimensional spectra have been averaged horizontally,
\begin{eqnarray}
  E_{\rm L}(k) &=& \frac{1}{N_x N_y}
                   \sum\limits_{p,q=1}^{N_x,N_y} |\tilde{u}_z(x_p,y_q,k)|^2 , \\
  E_{\rm T}(k) &=& \frac{1}{N_x N_y}
                   \sum\limits_{p,q=1}^{N_x,N_y}
                     \frac{1}{2} |\tilde{u}_\perp(x_p,y_q,k)|^2 ,
\end{eqnarray}
where $|\tilde{u}_\perp|^2=|\tilde{u}_x|^2+|\tilde{u}_y|^2$.
Here
\begin{equation}
  \tilde{\uu}(x,y,k) = \left(\frac{L_z}{2\pi}\right)^{1/2} \frac{1}{N_z}
               \sum\limits_{n=1}^{N_z} e^{ikz_n} \uu(x,y,z_n)
\end{equation}
is the one-dimensional Fourier transform of the velocity vector $\uu(\xx)$,
and $k$ scans the interval $[0,k_{\rm Ny}] = [0,\pi/\delta z]$ at the
resolution $\delta k = 2\pi/L_z$.
The three-dimensional spectrum $E(k)$ has been obtained by integrating
the three-dimensional spectral energy density over shells
$k_i-\delta k/2 \le k < k_i+\delta k/2 $ in analogy to
Eq.~(\ref{E-k-shellint}), with $\delta k = 2\pi/L_x$ (and $L_x=L_y=L_z$).
The spectra are normalized such that
\begin{eqnarray}
  \lefteqn{\int_0^\infty E(k)\,\dd k
           = \frac{1}{2}\langle\uu^2\rangle} \nonumber \\
  &\quad=& 3 \int_0^\infty E_{\rm L}(k)\,\dd k
     = 3 \int_0^\infty E_{\rm T}(k)\,\dd k .
\end{eqnarray}

\section{Relations between one- and three-dimensional spectra}
\label{S-1d-3d}

While most experimental measurements yield longitudinal
one-dimensional spectra $E_{\rm L}(k)$, the discussion of the relation between
one-dimensional and three-dimensional spectra is significantly simpler for
the total one-dimensional spectrum $E_{\rm 1D}(k)$.
We thus split this section in two parts: we first outline the relations
for the total one-dimensional spectrum $E_{\rm 1D}(k)$,
while in the second part
we obtain analogous results for longitudinal spectra, which are of
direct relevance for experiments.

\subsection{The total one-dimensional spectrum}
\label{S-E1}

The total one-dimensional spectrum $E_{\rm 1D}(k)$ is the sum of the
longitudinal and twice (for the two directions) the transversal
one-dimensional spectra,
\begin{equation}
  E_{\rm 1D}(k) = E_{\rm L}(k) + 2  E_{\rm T}(k) .
\end{equation}
It is thus in some sense `more isotropic' than its constituents which
results in simpler relations to the fully isotropic three-dimensional
spectrum.
In Appendix~\ref{Appendix-1D} we show that $E_{\rm 1D}(k)$ is related to
$E(k)$ by
\begin{equation}
  E_{\rm 1D}(k) = \int_k^\infty \frac{E(k')}{k'} \,\dd k' ,
  \label{E1D-from-E}
\end{equation}
which can also be inverted to give
\begin{equation}
  E(k) = -k\,\frac{\dd E_{\rm 1D}}{\dd k} \; ,
  \label{E-from-E1}
\end{equation}
provided the turbulence is isotropic.
As $E(k)$ must be positive,
Eq.~(\ref{E1D-from-E}) shows directly that the one-dimensional
spectrum $E_{\rm 1D}(k)$ must be monotonously decreasing.
On the other hand, no such restriction holds for $E(k)$, which is in fact
increasing near $k=0$, since $E(0)=0$.

Equation~(\ref{E-from-E1}) is a \emph{local} relationship, and
thus the functional form of $E(k)$ is fully determined by the local
behavior of $E_{\rm 1D}(k)$ at a given wave number.
To relate it to the bottleneck effect,
we introduce the compensated spectra
\begin{equation}
  \widetilde{E}_{\rm 1D}(k) \equiv k^{5/3} E_{\rm 1D}(k),\quad
  \widetilde{E}(k) \equiv k^{5/3} E(k),
  \label{compensated_def}
\end{equation}
and rewrite Eq.~(\ref{E-from-E1}) in the form
\begin{equation}
  \widetilde{E}(k)=
  \left[ \frac{5}{3} -\frac{\dd \ln \widetilde{E}_{\rm 1D}(k)}{\dd\ln k}
  \right]
  \widetilde{E}_{\rm 1D}(k).
  \label{E-from-E1-compensated}
\end{equation}
Several conclusions can be drawn from Eq.~(\ref{E-from-E1-compensated}).
First, if there is a finite interval where the Kolmogorov scaling
(\ref{Kolmogorov})
holds for the one-dimensional spectrum, then the same scaling will hold
for the three-dimensional spectrum, and vice versa.
Further, Eq.~(\ref{E-from-E1-compensated}) shows explicitly that even if
$\widetilde{E}_{\rm 1D}(k)$ is
monotonously decreasing, $\widetilde{E}(k)$ may show a maximum near
$k_{\rm d}$, provided that $\widetilde{E}_{\rm 1D}(k)$ bends towards the
dissipative range sufficiently suddenly.
For example, if the one-dimensional spectrum has the form
\begin{equation}
  E_{\rm 1D}(k)=k^{-5/3}\exp[-(k/k_{\rm d})^n],
  \label{exponential-1D}
\end{equation}
then for small wave numbers $k$ the compensated three-dimensional
spectrum behaves like
\begin{equation}
  \widetilde{E}(k)\approx\frac{5}{3}+\left(n-\frac{5}{3}\right)
  \left(\frac{k}{k_{\rm d}}\right)^n\quad\text{(for $k\ll k_{\rm d}$)},
\end{equation}
which shows that there will be a bottleneck effect in the three-dimensional
spectrum when the one-dimensional spectrum falls off with an exponent
$n>5/3$.
Finally, if $E_{\rm 1D}(k)$ shows a bottleneck effect,
i.e.~$\widetilde{E}_{\rm 1D}(k)$ shows a local maximum at some wave number
$k_{\rm m}$, then Eq.~(\ref{E-from-E1-compensated}) shows that
$\widetilde{E}(k)$ will
also have an enhanced value there,
$\widetilde{E}(k_{\rm m}) = (5/3)\widetilde{E}_{\rm 1D}(k_{\rm m})$,
and thus the
three-dimensional spectrum $E(k)$ must show a bottleneck effect, too.

\subsection{Longitudinal and transversal one-dimensional spectra}

For isotropic turbulence
the one-dimensional longitudinal and transversal spectra
are related to the three-dimensional spectrum by
(see Appendix~\ref{Appendix-longi-transv})
\begin{eqnarray}
  \label{EL-from-E}
  E_{\rm L}(k) &=& \frac{1}{2}
                    \int_k^\infty \left(1 - \frac{k^2}{k'^2}\right)
                                  \frac{E(k')}{k'} \,\dd k' , \\
  \label{ET-from-E}
  E_{\rm T}(k) &=& \frac{1}{4}
                    \int_k^\infty \left(1 + \frac{k^2}{k'^2}\right)
                                  \frac{E(k')}{k'} \,\dd k' ,
\end{eqnarray}
and
\begin{equation}
  \label{E-from-EL}
  E(k) = k^2 E_{\rm L}''(k) - k E_{\rm L}'(k) \\
\end{equation}
(the primes denoting derivatives).
Differentiating Eq.~(\ref{EL-from-E}), we obtain
\begin{equation}
  E_{\rm L}'(k) = - k \int_k^\infty \frac{E(k')}{k'^3} \, \dd k' <0 ,
\end{equation}
which shows that $E_{\rm L}(k)$ (just like $E_{\rm 1D}(k)$) must be
monotonously decreasing.

Equation~(\ref{E-from-EL}) is again a \emph{local} relationship, and
thus the functional form of $E(k)$ is fully determined by the local
behavior of $E_{\rm L}(k)$ at a given wave number.
As in \Eq{compensated_def}, we introduce the compensated spectra
\begin{equation}
  \widetilde{E}(k) \equiv k^{5/3} E(k), \quad
  \widetilde{E}_{\rm L}(k) \equiv k^{5/3} E_{\rm L}(k),
\end{equation}
and rewrite Eq.~(\ref{E-from-EL}) in the form
\begin{equation}
  \widetilde{E} =
     \frac{55}{9} \widetilde{E}_{\rm L}
   - \frac{13}{3} k \widetilde{E}_{\rm L}'
   + k^2 \widetilde{E}_{\rm L}'' .
  \label{E-from-EL-compensated}
\end{equation}
The conclusions that can be drawn from Eq.~(\ref{E-from-EL-compensated})
are similar to those for $E_{\rm 1D}(k)$.
If there is a finite interval where the Kolmogorov scaling
$E_{\rm L}(k) = C_{\rm K}^{\rm L} \epsilon^{2/3} k^{-5/3}$
holds for the longitudinal spectrum
(which implies $\widetilde{E}_{\rm L}' = \widetilde{E}_{\rm L}'' = 0$)
then the same scaling will hold for the three-dimensional spectrum, and
vice versa.
The Kolmogorov constant $C_{\rm K}$ of the three-dimensional spectrum
is then a factor $55/9\approx 6.1$ larger than the corresponding constant
$C_{\rm K}^{\rm L}$ for $E_{\rm L}(k)$; cf.\ Eq.~(16.37') of
Ref.~\cite{MoninYaglom}.

Equation~(\ref{E-from-EL-compensated}) shows that even if
$\widetilde{E}_{\rm L}(k)$ is
monotonously non-increasing, $\widetilde{E}(k)$ may show a maximum near
$k_{\rm d}$, provided that
the term involving $\widetilde{E}_{\rm L}'$ dominates over the curvature
term $k^2\widetilde{E}_{\rm L}''$.
Finally, if $E_{\rm L}(k)$ shows a bottleneck effect,
then beyond the local maximum, $\widetilde{E}_{\rm L}'(k)<0$.
If the maximum is wide enough, Eq.~(\ref{E-from-EL-compensated}) implies
that $\widetilde{E}$ will show a bottleneck effect, too.
The situation is less clear if
$k_{\rm m}^2 \widetilde{E}_{\rm L}''(k_{\rm m})$ is large (corresponding to a narrow
maximum of $\widetilde{E}_{\rm L}(k)$), but numerical experiments with model spectra suggest
that the three-dimensional spectrum $\widetilde{E}_{\rm L}(k)$ will still
be non-monotonous, although it may vary in a more complicated manner.

\section{Sample spectra}
\label{S-sample-specs}

\subsection{Model spectra}

\begin{figure}
  \centering
  \includegraphics[width=0.49\textwidth]{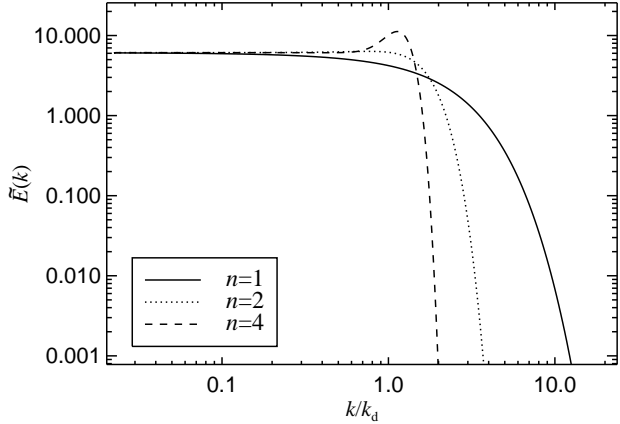}
  \caption{
    Compensated three-dimensional spectrum
    $\widetilde{E}(k) = (k/k_{\rm d})^{5/3} E(k)$
    corresponding to the longitudinal
    spectrum (\ref{exponential-L}) for exponents $n=1, 2, 4$.
    The bottleneck effect appears for $n \gtrsim 2$.
  }\label{Fexponential}
\end{figure}

For illustration purposes, we consider a longitudinal spectrum of the form
\begin{equation}
  E_{\rm L}(k) = \left(\frac{k}{k_{\rm d}}\right)^{-5/3}\exp[-(k/k_{\rm d})^n] .
  \label{exponential-L}
\end{equation}
The compensated spectrum
$\widetilde{E}_{\rm L}(k) \equiv (k/k_{\rm d})^{5/3} E_{\rm L}(k) = \exp[-(k/k_{\rm d})^n]$
is monotonously decreasing and thus $E_{\rm L}(k)$ shows no bottleneck
effect at all.
At $k=k_{\rm d}$ the compensated three-dimensional spectrum
has the value
\begin{equation}
  \widetilde{E}(k_{\rm d}) = e^{-1} \left( \frac{55}{9} + \frac{16}{3}n \right)
\end{equation}
which shows that $\widetilde{E}(k_{\rm d}) > 55/9$ (a sufficient
condition for the bottleneck effect in the three-dimensional spectrum)
if $n > (55/48)(e-1) \approx 1.97$; this is illustrated in \Fig{Fexponential}.
A more thorough analysis reveals that $\widetilde{E}(k)$ will have a
maximum if, and only if $n > 5/3$, but for $n\lesssim 2$ the maximum is
hardly discernible.

\begin{figure}
  \centering
  \includegraphics[width=0.49\textwidth]{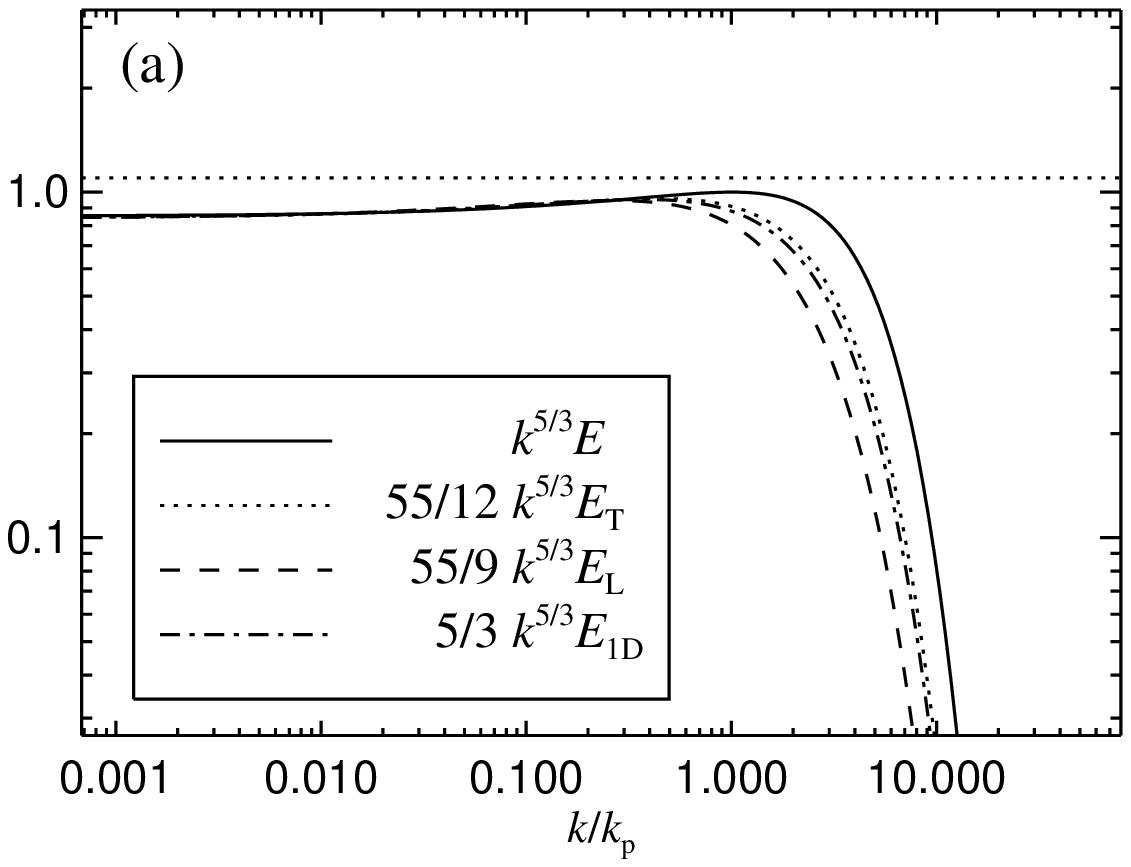}\\ 
  \includegraphics[width=0.49\textwidth]{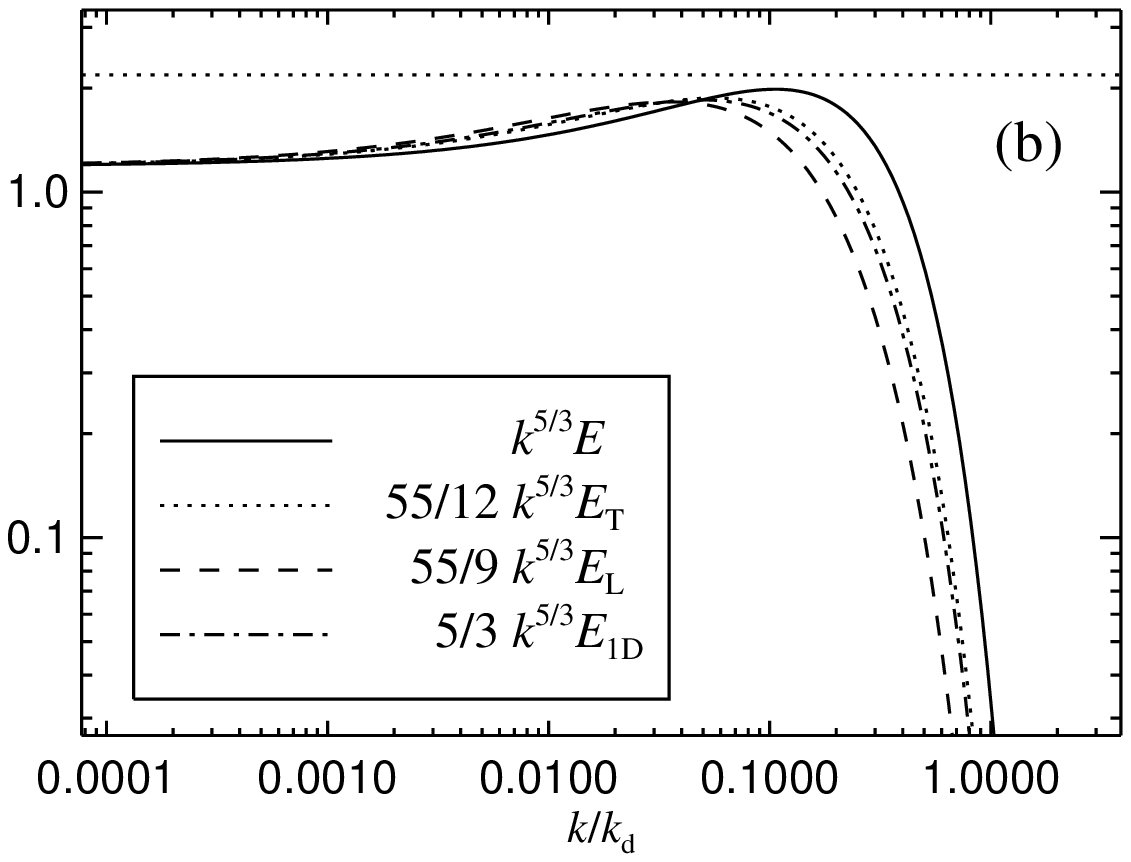}
  \caption{
    Comparison of compensated spectra $\widetilde{E}_{\rm L}(k)$ and
    $\widetilde{E}(k)$ for the models of She \& Jackson, and of Quian.
    (a) longitudinal spectrum (\ref{EL-She}), together with derived 
    three-dimensional spectra, longitudinal and total 1D spectra; the
    spectra are normalized according to
    $k_{\rm p} = E_{\rm L}(k_{\rm p}) = 1$.
    Note the appearance of a mild bottleneck effect, i.e.~a maximum in
    $\widetilde{E}(k)$.
    (b) Three-dimensional spectrum (\ref{E-Quian}), together with
    derived one-dimensional spectra;
    the spectra are normalized according to $k_{\rm d} = \epsilon = 1$.
    The bottleneck effect is quite pronounced and appears in both, one-
    and three-dimensional spectra.
    Note that in both plots
    the amplitudes of the one-dimensional spectra have been scaled to get
    matching plateaus in the inertial range.
    The dotted horizontal lines are drawn for orientation.
  }
  \label{tst_spec2}
\end{figure}

As we have seen,
in order to get a bottleneck effect, we need a one-dimensional spectrum
with a more complicated form than just $k^{-5/3}\exp(-k/k_{\rm d})$.
One functional form where this is given has been proposed by
She and Jackson \cite{SJ93}, based on experimental data:
\begin{equation}
  \frac{E_{\rm L}(k)}
       {E_{\rm L}(k_{\rm p})}
  = \left[ \left(\frac{k}{k_{\rm p}}\right)^{-5/3} \!\!
           + 0.8 \left(\frac{k}{k_{\rm p}}\right)^{-1}
    \right] \, e^{-0.63\,k/k_{\rm p}} \; .
\label{EL-She}
\end{equation}
Figure \ref{tst_spec2}(a) shows this spectrum [for
$E_{\rm L}(k_{\rm p}) = k_{\rm p} = 1$], together with the corresponding
three-dimensional spectrum (\ref{E-from-EL}).
The compensated spectra clearly show a bottleneck effect in the
three-dimensional spectrum, which is (practically) absent in the
one-dimensional spectrum.

Quian \cite{Quian1984} has proposed a spectrum that shows quite a marked
bottleneck effect.
Based on a closure model, he suggests the functional form
\begin{eqnarray}
  E(k) &=& \epsilon^{2/3} k^{-5/3}
           \left[1.19+6.31\left(\frac{k}{k_{\rm d}}\right)^{2/3}
           \right] \times \nonumber \\ 
       & & {} \times
           \exp[-5.4\,(k/k_{\rm d})^{4/3}] .
  \label{E-Quian}
\end{eqnarray}
Figure~\ref{tst_spec2}(b) shows the longitudinal and three-dimensional
spectra.
For this model, the bottleneck effect is very prominent in $E(k)$, and
also evident (although weaker) in the longitudinal one-dimensional
spectrum $E_{\rm L}(k)$.

\subsection{Spectra from direct numerical simulations}

\subsubsection{Three-dimensional spectrum from $E_{\rm L}(k)$}

%

\begin{figure}
  \centering
  \includegraphics[width=0.49\textwidth]{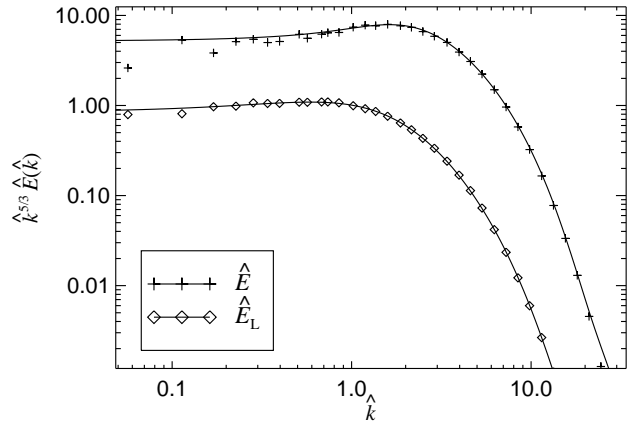}
  \caption{Compensated longitudinal energy spectrum
    $\hat{k}^{5/3}\hat{E}_{\rm L}(\hat{k})$ according to the
    parameterization~(\ref{Eq-param-512}) and corresponding
    three-dimensional spectrum $\hat{k}^{5/3}\hat{E}(\hat{k})$ from
    Eq.~(\ref{E-from-EL}).
    For comparison, crosses and diamonds show the spectra obtained
    directly from the simulation.
    The spectra have been normalized by introducing
    $\hat{k} \equiv k/k_{\rm p}$,
    $\hat{E}_{\rm L} \equiv E_{\rm L} / E_{\rm L}(k_{\rm p})$ and
    $\hat{E} \equiv E / E_{\rm L}(k_{\rm p})$, where $k_{\rm p}$ is the
    peak dissipation wavenumber.
    The three-dimensional energy spectrum agrees quite well with the
    numerical data for $k \ge 4\,\delta k$, where
    $\delta k = 2\pi/L_x \approx 0.06\,k_{\rm p}$ is the wave number
    resolution.
  }\label{parameterized-512}
\end{figure}

To avoid (double) numerical differentiation of our spectra, we use a
parameterization for $E_{\rm L}(k)$.
The one-dimensional spectrum shown in Fig.~\ref{pspec_bottleneck_512} is
well approximated by the formula
\begin{equation}
  E_{\rm L}^{(\rm p)}(k)
  = \bigl(\hat{k}^{-5/3} + a_0 \hat{k}^{a_1}\bigr)
        e^{-a_2 \hat{k}} \;
        \frac{a_3 + a_4 \hat{k} + a_5\hat{k}^2}
             {1 + a_6 \hat{k} + a_7\hat{k}^2} \; ,
  \label{Eq-param-512}
\end{equation}
with
$a_i = (
2.1{\times}10^{-8},
4.3,
0.42,
0.85,
1.2,
0.00048,\allowbreak
-0.0068,\allowbreak
0.34)$,
where $\hat{k} \equiv k/k_{\rm p}$.
The peak dissipation wavenumber $k_{\rm p} \approx 18$ is the
location of the maximum of the dissipation spectrum $k^2 E_{\rm L}$, and is
about one order of magnitude smaller than $k_{\rm d}$ \cite{SJ93}.
Figure~\ref{parameterized-512} shows the parameterized one-dimensional
spectrum $E_{\rm L}(k)$ according to Eq.~(\ref{Eq-param-512}),
together with the derived three-dimensional spectrum $E(k)$ from
Eq.~(\ref{E-from-EL}).
Also shown are data points from the numerical simulation (diamonds and
crosses).
Comparing the calculated three-dimensional profile (solid line) with the
data points (crosses), we find that Eq.~(\ref{E-from-EL}) agrees quite
well with the numerical data for not too small $k$.
The discrepancy for very small wave numbers can be explained by the fact
that the periodicity of the numerical box
precludes isotropy at the largest scales.

It is quite evident from Fig.~\ref{parameterized-512} that
$\widetilde{E}(k)$ shows a
much more pronounced local maximum near the dissipation wavenumber
$k_{\rm d}$ than does $\widetilde{E}_{\rm L}(k)$.
The width and structure of the maximum is well-reproduced by
Eq.~(\ref{E-from-EL}) applied to our parameterization (\ref{Eq-param-512})
of $E_{\rm L}(k)$.

\subsubsection{One-dimensional spectra from $E(k)$}

\begin{figure}
  \centering
  \includegraphics[width=0.49\textwidth]{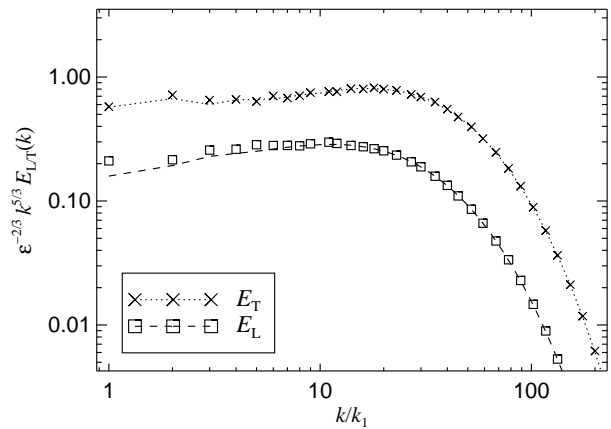}
  \caption{Compensated one-dimensional energy spectra.
    Shown are
    the longitudinal spectrum $E_{\rm L}(k)$ and the transversal spectrum
    $E_{\rm T}(k)$.
    Data points obtained directly from the numerical simulation are
    indicated as boxes ($E_{\rm L}$) and crosses ($E_{\rm T}$).
    The dashed and dotted lines show the one-dimensional
    spectra obtained from Eqs.~(\ref{EL-from-E}) and
    (\ref{ET-from-E}), respectively.
    The spectra agree quite well with the
    numerical data for $k \ge 2\,\delta k$.
  }
  \label{1d-long-transv}
\end{figure}

In contrast to experiments, data from numerical simulations easily provide
the three-dimensional spectrum $E(k)$.
Sometimes it may be interesting to determine the one-dimensional spectra
$E_{\rm L}(k)$ and $E_{\rm T}(k)$ from the three-dimensional spectrum.
The corresponding relations (\ref{EL-from-E}), (\ref{ET-from-E}) were
given above.
Here we apply them to our numerical data to see how well the inferred
one-dimensional spectra agree with those directly obtained from the
simulation data.
Figure~\ref{1d-long-transv} shows the longitudinal and transversal
spectra obtained by applying Eqs.~(\ref{EL-from-E}) and (\ref{ET-from-E})
to the three-dimensional spectrum using the trapezoidal rule.
The agreement is quite good for both one-dimensional spectra apart from
the very lowest wave number.

\section{Structure functions}

Another classical tool in turbulence research is the investigation of structure
functions \cite{LohseMuellerG}. The second-order structure function,
\begin{equation}
  S_2(r) = \left<[\uu(\xx) - \uu(\xx{+}\rr)]^2
               \right>,
\end{equation}
is related to the three- and one-dimensional
spectra via the Fourier-type integral transforms \cite[cf.][]{MoninYaglom}
\begin{eqnarray}
    S_2(r) &=& 4 \int\limits_0^\infty
                   \left( 1 - \frac{\sin kr}{kr} \right)
                    E(k) \, dk \\
           &=& 4 \int\limits_0^\infty
                   \bigl( 3 - 3\cos kr + kr \sin kr \bigr)
                    E_{\rm L}(k)\, dk \; . \quad
\end{eqnarray}
Similarly, the longitudinal and transversal second-order structure
functions
\begin{eqnarray}
  S_2^{\rm(L)}(r) &=& \left<[u_z(\xx) - u_z(\xx{+}r\hat{\bm{z}})]^2
                      \right>, \\
  S_2^{\rm(T)}(r) &=& \left<[u_z(\xx) - u_z(\xx{+}r\hat{\bm{x}})]^2
                      \right>,
\end{eqnarray}
can be expressed as
\begin{eqnarray}
    S_2^{\rm(L/T)}(r)
    = 4 \int\limits_0^\infty
                   \left( 1 - \cos kr \right)
                    E_{\rm L/T}(k) \, dk .
\end{eqnarray}

Localized variations of $E(k)$ or $E_{\rm L}(k)$ in wavenumber space, such
as the bottleneck effect, will influence
$S_2(r)$ and $S_2^{\rm(L/T)}(r)$ in a nonlocal fashion.
Correspondingly, little insight into the bottleneck effect can be
expected from structure functions.
On the other hand, this means that structure functions are less sensitive
to the bottleneck
effect and might thus be a more robust tool for assessing
scaling exponents and possibly even the Kolmogorov constant at moderate
Reynolds number.

However, while structure functions are much smoother than spectra, their
scaling range is considerably smaller, which adds its own difficulties
to that method.
In three numerical simulations at $256^3$, $512^3$ and $1024^3$ grid
points, we find the value of the structure function exponent derived from
$S_2(r)$ to be $0.74$, $0.67$, and $0.68$, respectively, which is quite
close to Kolmogorov's value of $2/3$.
However, if we use the the two transversal and the longitudinal structure
function (corresponding to the $z$-displacement of the $u_x$, $u_y$ and
$u_z$ component of the velocity vector), the values are far less accurate
and span the intervals
$[0.67,0.79]$, $[0.62,0.72]$, $[0.60,0.72]$, respectively, for the three
resolutions, which indicates that the convergence is maybe not
yet reached.

The situation for the Kolmogorov constant is even worse, because from
$S_2(r)$ we get the values $3.96$, $1.93$, $1.8$, respectively, for the
three resolutions.
So in practice, the slow convergence may well render this method more
unreliable than the use of one-dimensional spectra.

\bigskip

\begin{figure}
  \centering
  \includegraphics[width=0.48\textwidth]{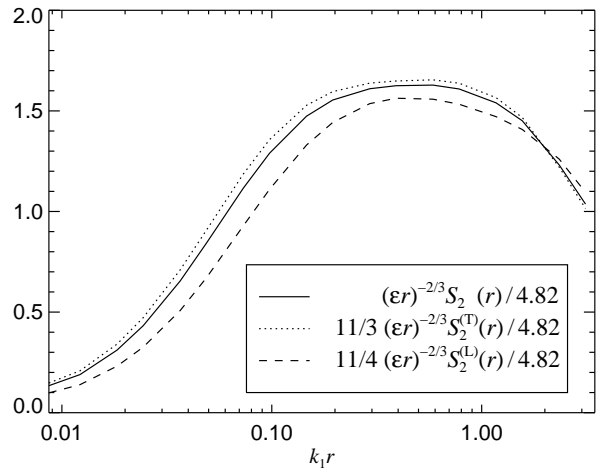}
  \caption{Compensated second-order structure functions from our direct
    numerical simulation data.
    The amplitudes have been scaled to get matching plateaus representing
    the Kolmogorov constant $C_{\rm K}$.
  }
  \label{F-structfunc}
\end{figure}

Similar to the spectra, the second-order structure functions can be
transformed into each other (assuming isotropy) according to the relations
\begin{eqnarray}
  \label{S2-from-S2L}
  S_2(r)       &=& r \frac{dS_2^{\rm(L)}(r)}{dr} + 3 S_2^{\rm(L)}(r)     ,\\
  \label{S2L-from-S2}
  S_2^{\rm(L)}(r) &=& \frac{1}{r^3} \int_0^r S_2(r')\,r'^2\;dr' ,\\
  \label{S2T-from-S2-S2L}
  S_2^{\rm(T)}(r) &=& \frac{S_2(r) - S_2^{\rm(L)}(r)}{2} \; .
\end{eqnarray}
Like in the case of relations (\ref{EL-from-E})--(\ref{E-from-EL}), this
has implications for the monotonicity of the compensated structure
functions $\widetilde{S}(r) \equiv S r^{-2/3}(r)$ near the boundaries of
the scaling interval; see also Ref.~\cite{LohseMuellerG}.
Normally, however, none of the structure functions $S_2(r)$,
$S_2^{\rm(L)}(r)$, $S_2^{\rm(T)}(r)$ show a secondary bump near the edges of the
scaling interval, as can be seen in Fig.~\ref{F-structfunc}, which implies
that we cannot directly apply the results of Secs.~\ref{S-1d-3d} and
Sec~\ref{S-sample-specs} to structure functions.

\section{Conclusions}

In this paper we have highlighted the discrepancy between one-dimensional
and three-dimensional spectra.
Well-known relations between the spectra show that the three-dimensional
spectrum may show the bottleneck effect even if the one-dimensional
spectra do not show it at all, while the converse cannot happen.
The spectra always agree in the inertial range,
$E(k) \propto E_{\rm L}(k) \propto k^{-5/3}$, but in current numerical
simulations the length of the inertial range is limited to about one
decade, so the discrepancy is quite noticeable.
Indeed, the topic of a bottleneck effect in hydrodynamic turbulence
has only emerged in the last ten years since numerical simulations have
shown this to be a strong effect.
On the other hand, the relation linking one-dimensional
to three-dimensional spectra has been known for fifty years, but it
has to our knowledge never been explicitly discussed in connection with the
bottleneck effect.
In the present paper we have shown that much of the bottleneck effect
seen in numerical turbulence simulations is simply the result of the
mathematical discrepancy between one- and three-dimensional spectra.

The bottleneck effect has no evident manifestation in the second-order
structure functions, where localized features in $k$-space
appear in a delocalized manner.
This implies that in order to obtain the asymptotic energy spectrum
exponent it may be easier to use second-order structure function
exponents, although in practice the reduced scaling range may render this
method difficult.


\appendix*

\section{Relations between one- and three-dimensional spectra}

In this Appendix, we derive the relations between the three-dimensional
spectrum $E(k)$ and the one-dimensional spectra $E_{\rm L}(k)$, $E_{\rm L}(k)$,
$E_{\rm 1D}(k)$, most of which can also be
found in textbooks like Refs.~\cite{Batchelor,Hinze,MoninYaglom}.

\subsection{Total one-dimensional spectrum}
\label{Appendix-1D}
We derive the relation between the
three-dimensional spectrum $E(k)$ and the total one-dimensional spectrum
$E_{\rm 1D}(k) \equiv E_{\rm L}(k) + 2 E_{\rm T}(k)$.
Consider a periodic box of volume $V = L_x L_y L_z$ with a turbulent
velocity field $\uu(\xx)$, which has the Fourier transform
\begin{equation}
  \hat{\uu}(\kk) = \frac{1}{\sqrt{(2\pi)^3 V}}
             \int_{V} e^{i\kk\cdot\xx} \uu(\xx)\, \dd x^3 ,
\end{equation}
with the inversion
\begin{equation}
  \uu(\xx) = \sqrt{\frac{V}{(2\pi)^3}}
             \int e^{-i\kk\cdot\xx} \hat{\uu}(\kk)\, \dd k^3 .
\end{equation}
The one-dimensional kinetic energy spectrum is
\begin{equation}
  E_{\rm 1D}(k_z) =
  2 \int\!\!\!\int \frac{\left<|\hat{\uu}(\kk)|^2\right>}{2}
      \, \dd k_x\,\dd k_y ,
  \qquad(k_z \ge 0) ,
  \label{E1-def}
\end{equation}
where $\left<\cdot\right>$ denotes an ensemble average, and
$\kk = (k_x,k_y,k_z)$.
The factor $2$ in Eq.~(\ref{E1-def}) accounts for the fact that
$E_{\rm 1D}$ does 
not distinguish between positive and negative $k_z$.
Normalization of $E_{\rm 1D}(k_z)$ is such that
\begin{equation}
  \int_0^\infty E_{\rm 1D}(k_z) \,\dd k_z = \frac{u_{\rm rms}^2}{2}
  \equiv \frac{1}{V} \int_V \frac{\left<|\uu(\xx)|^2\right>}{2} \, \dd x^3.
\end{equation}
Equation (\ref{E1-def}) can also be written as the $xy$-average
\begin{equation}
  E_{\rm 1D}(k_z) = \frac{1}{L_x L_y}\int \left<|\tilde{\uu}(x,y,k_z)|^2\right>
                                \,\dd x\, \dd y
\end{equation}
and is for homogeneous turbulence equal to
$\left<|\tilde{\uu}(x,y,k_z)|^2\right>$ at any point $(x,y)$.

The three-dimen\-sional velocity energy spectrum is given by
\begin{equation}
  E(k) \equiv
  \int_{4\pi} \frac{\left<|\hat{\uu}(\kk)|^2\right>}{2} k^2 \,\dd\Omega_k ,
  \label{E-k-shellint}
\end{equation}
where $\dd\Omega_k$ denotes the solid angle element in $\kk$-space.
$E(k)$ satisfies the relation
\begin{equation}
  \int_0^\infty E(k) \, \dd k = \frac{u_{\rm rms}^2}{2} .
\end{equation}

If $\uu$ is statistically isotropic in the sense that the ensemble average
of the spectral energy of the velocity $\left<|\uu(\kk)|^2\right>$ is only
a function of $k = |\kk|$, then $E(k)$ becomes
\begin{equation}
  E(k) = 4\pi k^2 \frac{\left<|\hat{\uu}(\kk)|^2\right>}{2} .
  \label{E-uu-kk}
\end{equation}
To evaluate $E_{\rm 1D}$ in this case, we introduce cylindrical coordinates
$(\kappa,\phi,k_z)$ in $\kk$-space and write the double integral
(\ref{E1-def}) in the form
\begin{eqnarray}
  E_{\rm 1D}(k_z)
  &=& 2 \int_{0}^{\infty} \frac{\left<|\hat{\uu}(\kk)|^2\right>}{2} \,
                          2\pi\kappa\, \dd\kappa \nonumber \\
  &=& 4\pi \int_{k_z}^{\infty}
                          \frac{\left<|\hat{\uu}(\kk)|^2\right>}{2}\,
                          k\, \dd k ,
  \label{E1-kappa-k}
\end{eqnarray}
since $\kappa^2 = k^2 - k_z^2$.
Comparing with Eq.~(\ref{E-uu-kk}), we see that
\begin{equation}
  E_{\rm 1D}(k_z) = \int_{k_z}^{\infty} \frac{E(k)}{k} \, \dd k ,
\end{equation}
the inversion of which gives
\begin{equation}
  E(k) = - k \frac{\dd E_{\rm 1D}(k)}{\dd k}
       = - E_{\rm 1D} \frac{\dd\ln E_{\rm 1D}(k)}{\dd\ln k} \; .
\end{equation}

\subsection{Longitudinal and transversal one-dimensional spectra}
\label{Appendix-longi-transv}
In this section we derive the relation between longitudinal and
transversal one-dimensional energy spectra $E_{\rm L}(k)$,
$E_{\rm T}(k)$, and the three-dimensional energy spectrum $E(k)$.

For homogeneous, isotropic turbulence, the energy spectrum tensor is given by
\cite[see, e.g.,][]{MoninYaglom,Batchelor,Hinze}
\begin{eqnarray}
  F_{pq}(\kk)
  &\equiv& \left<\hat{u}_p(\kk)\hat{u}_q^*(\kk)\right> \nonumber \\
  &=& \left[ F_{\rm L}(k) {-} F_{\rm T}(k) \right]
      \frac{k_p k_q}{k^2}
      + F_{\rm T}(k)\, \delta_{pq} \; . \quad
\end{eqnarray}
If we assume incompressibility, the longitudinal component
$F_{\rm L}$ vanishes, and thus
\begin{equation}
  F_{pq}(k)
  = \left( \delta_{pq} - \frac{k_p k_q}{k^2} \right)
    F_{\rm T}(k)
  = \left( \delta_{pq} - \frac{k_p k_q}{k^2} \right)
    \frac{E(k)}{4\pi k^2} \; ,
\end{equation}
and in particular
\begin{eqnarray}
  2\pi F_{zz}(k)
  &=& \left( 1 - \frac{k_z^2}{k^2} \right) \frac{E(k)}{2\,k^2} , \\
  2\pi \bigl[F_{xx}(k)+F_{yy}(k)\bigr]
  &=& 
  \left( 1 + \frac{k_z^2}{k^2} \right) \frac{E(k)}{2\,k^2} .
\end{eqnarray}

The longitudinal one-dimensional spectrum
\begin{equation}
  E_{\rm L}(k_z)
  \equiv
  2 \int\!\!\!\int \frac{F_{zz}(\kk)}{2} \, \dd k_x\,\dd k_y ,
\end{equation}
thus becomes
\begin{equation}
  E_{\rm L}(k_z)
  = \int\limits_0^\infty F_{zz}(k)\, 2\pi \kappa \, \dd\kappa
  = \frac{1}{2}
    \int\limits_{k_z}^\infty
      \left( 1 - \frac{k_z^2}{k^2} \right) \frac{E(k)}{k} \, \dd k ,
  \label{EL-from-E-App}
\end{equation}
using the same substitution as in Eq.~(\ref{E1-kappa-k}) above.
Similarly, we can write the transversal one-dimensional spectrum
\begin{equation}
  E_{\rm T}(k_z)
  \equiv
  2 \int\!\!\!\int \frac{F_{xx}(\kk){+}F_{yy}(\kk)}{4} \, \dd k_x\,\dd k_y ,
\end{equation}
in the form
\begin{equation} \label{ET-from-E-App}
  E_{\rm T}(k_z)
  = \frac{1}{4}
      \int_{k_z}^\infty
        \left( 1 + \frac{k_z^2}{k^2} \right) \frac{E(k)}{k} \, \dd k .
\end{equation}

Taking the derivative of Eq.~(\ref{EL-from-E-App}), we find
\begin{equation}
  \frac{E_{\rm L}'(k_z)}{k_z}
  = -\int_{k_z}^\infty \frac{E(k)}{k^3} \, \dd k ,
\end{equation}
and thus
\begin{equation}
  E(k) = k^2 E_{\rm L}''(k) - k E_{\rm L}'(k) .
\end{equation}
Inserting this relation into Eq.~(\ref{ET-from-E-App}) allows us to express
$E_{\rm T}(k)$ through $E_{\rm L}(k)$ as
\begin{equation}
  E_{\rm T}(k) = - k \frac{E_{\rm L}'(k)}{2} + \frac{E_{\rm L}(k)}{2} .
\end{equation}

\begin{acknowledgments}
We thank {\AA}ke Nordlund for useful comments on the subject.
Use of the parallel computers in Trondheim (Gridur), Odense (Horseshoe)
and Leicester (Ukaff) is acknowledged.
\end{acknowledgments}



\end{document}